# New route for stabilization of 1T-WS$_2$ and MoS$_2$ phases

Andrey N. Enyashin,*[a,b] Lena Yadgarov,[c] Lothar Houben,[d] Igor Popov,[e] Marc Weidenbach,[d] Reshef Tenne,[c] Maya Bar-Sadan*[d], and Gotthard Seifert[a]

The phenomenon of a partial 2H→1T phase transition within multiwalled WS$_2$ nanotubes under substitutional Rhenium doping is discovered by means of high-resolution transmission electron microscopy. Using density-functional calculations for the related MoS$_2$ compound we consider a possible origin of this phase transition, which was known formerly only for WS$_2$ and MoS$_2$ intercalated by alkali metals. An interplay between the stability of layered or nanotubular forms of 2H and 1T allotropes is found to be intimately related with their electronic structures and electro-donating ability of an impurity.

## 1. Introduction

Co-generic molybdenum and tungsten disulfides (MS$_2$, M = Mo, W) are typical representative compounds of the rather extensive family of layered chalcogenides. Their metal atoms have a six-fold coordination environment and are hexagonally packed between two trigonal atomic layers of S atoms. Depending on the arrangement of the S atoms two kinds of the hexagonal S-M-S triple layers are possible, which are composed by either prismatic $D_{3h}$- or octahedral $O_h$-MS$_6$ units (Fig. 1). Such triple layers interact by weak van-der-Waals interactions and may be stacked in different ways. For example, natural MoS$_2$ occurs as a mixture of two stable polymorphs based on $D_{3h}$-MoS$_6$ units: the hexagonal 2H-MoS$_2$ and the rhombohedral modification 3R-MoS$_2$, which unit cells include two and three S-M-S monolayers, respectively [1]. The 2H-polytype is dominant and more stable, hence the 3R-polytype transforms into the first one upon heating. Further polytypic structures, other than 2H or 3R, have been predicted [2], but have not yet been observed experimentally.

A MS$_2$ polytype based on $O_h$-MS$_6$ units has not been found in the nature and it is recognized as unstable [1]. Nevertheless, S-M-S layers with an octahedral coordination of metal atoms can be synthesized under intercalation of a 2H-MS$_2$ host lattice by alkali-metals Li and K, forming 1T-Li$_x$MS$_2$ and 1T-K$_x$MS$_2$ phases [3-5]. Moreover, subsequent solvation and reduction of these intercalates leads to the exfoliation and release of free 1T-MS$_2$ layers [5-8]. Depending on the composition of the intercalate, reducing agent used and temperature, the monolayers of 1T-MoS$_2$ and 1T-WS$_2$ can restack to form various superstructures as revealed by Raman-scattering [9], electron and x-ray diffraction [9-11] and scanning tunneling microscopy [12]. However, these metastable materials undergo an irreversible transition to 2H-MS$_2$ phases already at 95 °C [5]. Theoretical considerations of the interlayer S/S and intra-layer M/S glide within MoS$_2$ crystals clearly explained the preference of the 2H-MoS$_2$ allotrope over the 1T-MoS$_2$ structure [13].

Nowadays, micro- and nanosized particles of 2H-MoS$_2$ and 2H-WS$_2$ play a major role in the catalytic refinement of petroleum oils [14]. Numerous quantum-mechanical calculations reveal a correlation between their catalytic activity and metallic character of their edges [15-17], while the bulk crystals of these compounds are semiconductors [18,19]. Interestingly, quantum-mechanical calculations for the band structure of the unstable 1T-MoS$_2$ suggest that the ground state of this system is metallic and consequently it may possess high catalytic activity [20,21]. Thereby, there is an interest to search for a possible way of stabilizing the 1T-MoS$_2$ and 1T-WS$_2$ phases. Recently, significant progress has been achieved in doping MoS$_2$ and WS$_2$ closed-cage structures by Re [22], rendering the structures superior lubricating properties.

In this work we report the discovery of 1T-MS$_2$ phase within WS$_2$ nanotubes substitutionally doped by Rhenium. Such doped and still covalently bound lattice of the 1T-phase should be genuinely more stable comparing with the lattice of alkali-intercalated compound. Using quantum-mechanical calculations for the related MoS$_2$ compound we consider a possible origin of this phase and its stability in layered and nanotubular states, which are found to be intimately related with the electronic structures of 2H- and 1T-MS$_2$ allotropes and electro-donating character of Re impurities. Finally, we hypothesize about high catalytic and tribologic characteristics of these doped nanotubes and fullerene-like particles.

## 2. Methods

### 2.1. Synthesis of Re (~2 at%) doped WS$_2$ nanotubes

ReO$_3$ was added to a vertical quartz ampoule at its "hot" zone (bottom of the ampoule) containing inorganic WS$_2$ nanotubes at the "cold" zone (middle of the ampoule) and I$_2$ (~80 mg), which later serves as carrier agent (see Fig. 1). The ampoule was cooled with liquid nitrogen, evacuated to ~$10^{-5}$ Torr and sealed. The distance between the hot and the cold zone was ~12 cm, long enough to avoid excess doping. The ampoule was placed in a two-zone furnace and allowed to react for 22 hours at 800$^0$C (cold zone) to 950$^0$C (hot zone). The temperature gradient is intended to prevent back transport of the product. Afterwards the ampoule was quenched in order to stop the reaction at once. The rhenium composition $x$ was estimated by EDS (energy dispersive X-ray spectroscopy), and analyzed by SEM (scanning electron microscopy) and TEM (transmission electron microscope).

## 2.2. Microscopy Methods

Electron microscopy images were taken with an image-side aberration-corrected FEI Titan 80-300 transmission electron microscope operated at 300 kV. Negative spherical aberration imaging (NCSI) conditions were applied [23,24]. These conditions produce high sensitivity even for light elements and yield images that are a close representation of the projected potential for sufficiently thin samples, meaning that the image intensity can be correlated with the location of the atoms Phase retrieval from focal series was employed to enhance the signal-to-noise ratio and to eliminate the effect of residual aberrations [25].

Image simulations for NCSI conditions were done using model nanostructures with multislice calculations, to confirm the coordination between Mo and S atoms. Special care was taken to ensure, that both pattern and spatial distances match the models unequivocally, since different projections and tilts of the model might yield similar patterns.

## 2.3. Models and Computational Methods

Supercells of $MoS_2$ monolayers, double layers, single- and double-walled nanotubes were chosen as the models for the study of pure and doped 1T- and 2H-allotropes. $MoS_2$ monolayer and double layer supercells were composed of 10x10x1 and 5x5x2 $MoS_2$ unit cells, respectively. The models of $MoS_2$ nanotubes were represented with combinations of supercells for (14,14) and (21,21) nanotubes each composed of 7 elementary unit cells. The classification and construction principles of single-walled $MoS_2$ nanotubes was reported earlier [26].

All of the stability and electronic structure calculations with full geometry optimization and molecular dynamic simulations were performed with the density-functional-based tight-binding (DFTB) method [27,28], which formerly was applied in numerous studies of $MoS_2$ nanostructures [26,29,30].

## 3. Results and Discussion

### 3.1. Microscopy data

The direct observation by electron microscopy of Re atoms within the lattice of $MoS_2$ or $WS_2$ is practically impossible because of several reasons: the doping level of the nanotubes is very low (<1%) and the doping is homogeneous, therefore, the Re atoms do not produce much disorder within the lattice. The atomic number of Re is close to Mo and much smaller than W, therefore it does not create significant change of contrast in the images. The low doping level also hinders the use of electron energy loss spectroscopy (EELS) to detect the existence and place of the single Re atoms. However, EXAFS experimental results confirm the existence of individual atoms of Re as well as very small clusters of them within the lattice [22].

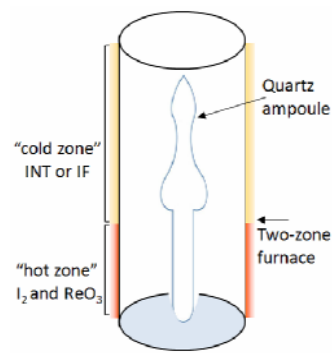

Fig. 1. Schematic drawing of the experimental setup: the reactants ($WS_2$ nanotubes, $ReO_3$ and $I_2$) are put within an evacuated quartz ampoule placed inside a two-zone furnace.

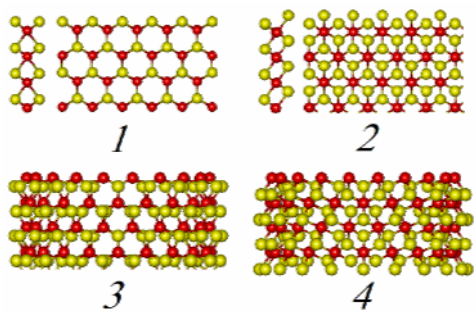

Fig. 2. Monolayers of 2H- (1) and 1T-allotropes (2) of transition metal sulfides $MS_2$ (M = W, Mo) composed of prismatic or octahedral units $MS_6$, respectively (the side and the top views are depicted). Below the models of single-walled zigzag (20,0) nanotubes are shown for 2H- (3) and 1T-allotropes (4) (lateral view).

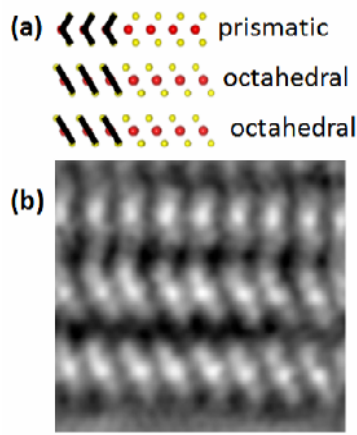

Fig. 3. A schematic representation of the octahedral and the prismatic polytypes of $MoS_2/WS_2$ (a). The prismatic layer is imaged as a chevron pattern, while the octahedral one creates a diagonal pattern. Mo/W in red, S in yellow. (b) Wavefunction reconstructed from a focal series taken at 300 kV, showing the outermost walls of Re-doped $WS_2$ nanotubes. As it is presented in (a), the two outermost layers are octahedral and the inner one is prismatic.

The different structural phases of $WS_2$ are directly distinguishable by HRTEM. The prismatic coordination of the S-M-S creates a chevron pattern, while the octahedral one is imaged as a diagonal line (See Fig. 3a) [30]. In the nanotubes presented, two layers of 1T-phase with octahedral coordination of metal atoms are seen at the outermost shells, followed by a prismatic inner layer of 2H-phase (see Fig. 3b). The observation of a few dozens of nanotubes showed a dominance of prismatic $D_{3h}$-$MS_6$ coordination while patches of single octahedrally coordinated layers can be found at or close to the nanotube surface at 10% of the structures..

In further, we employ the density-functional calculations in order to explain the occurence of unstable 1T-$MS_2$ phase.

## 3.2. Stability of undoped 1T-$MoS_2$ nanotubes

The results of DFTB calculations for the stability of pure 1T-$MoS_2$ single-walled nanotubes are shown in Fig. 4. It can be seen, that the curve of strain energy ($E_{str}$) versus radius, relative to the monolayer, shows similar trends in analogy to many other compounds, e.g. carbon and 2H-$MoS_2$ [26,31]. The strain energy is inversely proportional to the radius $R$ as $E_{str} = a/R^2$ and is also nearly independent of the chirality type. At the same time, the energies indicate a much lower stability of 1T-$MoS_2$ nanotubes comparing to the nanotubes based on the 2H-$MoS_2$ allotrope due to both the higher energy of 1T-monolayer on 0.27 eV/atom and a slightly larger strain energy factor $a = 18.6$ eV·Å$^2$/atom versus 15.8 eV·Å$^2$/atom for 2H-$MoS_2$ nanotubes.

The calculated high energies of 1T-$MoS_2$ nanotubes might hint that the much more stable inner core of 2H-$WS_2$ nanotubes stabilizes the 1T-$WS_2$ phase observed in the HRTEM of multiwalled sulfide nanotubes. In order to prove this idea, additional molecular dynamics simulations of double-walled $MoS_2$ nanotubes composed by single-walled nanotubes of 1T- and 2H-$MoS_2$ modifications with all possible mutual arrangements were performed (Fig. 5). However, according to the calculations, double-walled nanotubes containing at least one shell of 1T-$MoS_2$ phase undergo essential distortions or even disintegrate into a bundle of nanostripes. Thus, it is likely that the formation of the unfavorable 1T-phase within a multiwalled sulfide nanotubes is enhanced by the influence of Re in Re-doped $MS_2$ nanotubes.

Calculated densities-of-states (DOS) for some selected 1T- and 2H-$MoS_2$ nanostructures are drawn on Fig. 6. In both cases of allotropes there is no essential difference in the DOS profiles between the nanotubes of different chiralities and for the corresponding bulk and monolayer.

In the calculated electronic structure of the most stable 2H-$MoS_2$ allotrope three bands can be clearly distinguished, which agrees with the results of former experimental and quantum-mechanical investigations [19]. The valence band composed mainly of the S$3p$-states is separated from the Fermi level by ~3.5 eV. The states below and near the Fermi level are Mo$4d$-states. The bottom of the conduction band is also composed of Mo$4d$-states. Therefore, 2H-$MoS_2$ is a semiconductor with a band gap around 1.2 – 1.3 eV, which in the case of nanotubes may be slightly smaller depending on their radii [26]. In the framework of crystal field theory the semiconducting nature of 2H-$MoS_2$ is caused by the symmetry-induced splitting of the Mo$4d$-orbitals of a $D_{3h}$-$MoS_6$ unit into three groups: (1) Mo$4d_{z^2}$, which is completely occupied, (2) Mo$4d_{xy}$ and Mo$4d_{x^2-y^2}$ and (3) Mo$4d_{xz}$ and Mo$4d_{yz}$, which are unoccupied (Fig. 6).

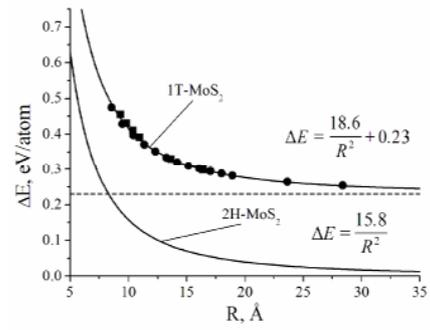

Fig. 4. The energies $\Delta E$ of 1T-$MoS_2$ monolayer (dashed line) and 1T-$MoS_2$ based single-walled nanotubes (■ – *zigzag* and ● – *armchair* chiralities) relative to the energy of 2H-$MoS_2$ based monolayer. For comparison the energies of 2H-$MoS_2$ single-walled nanotubes are also depicted (adapted from [26]).

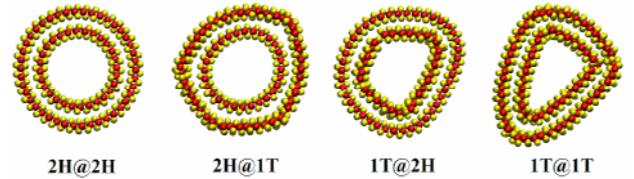

Fig. 5. Final MD snapshots (T = 300 K) for double-walled (14,14)@(21,21)$MoS_2$ nanotubes composed by the single-walled nanotubes of different allotropes: the layers with prismatic (like in 2H-phase) and octahedral (like in 1T-phase) coordination of metal atoms. All nanotubes containing 1T modification are unstable and decompose partially or totally into a bundle of nanostripes.

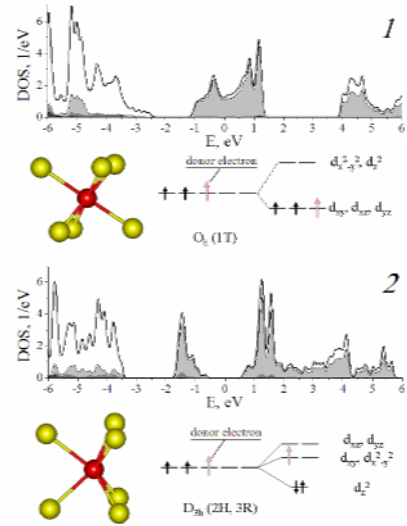

Fig. 6. Total (full line) and partial Mo$4d$ densities of states (DOS) for (14,14) 1T-$MoS_2$ (*1*) and - (14,14) 2H-$MoS_2$ nanotubes (*2*) and their simplified representations within crystal field theory. Mo$4d$-states are painted in gray. Fermi level is set to 0.0 eV. An electron donating of Mo$d_{xy}$ levels can stabilize 1T-$MoS_2$ phase and destabilize 2H-$MoS_2$ phase.

Both the monolayer and the nanotubes of 1T-$MoS_2$ allotrope show metal-like character (see Fig. 6). It agrees with the results of previous calculations and experimental data [20,21]. Like in the case of 2H-$MoS_2$ the valence band of the S$3p$-states is separated from the Fermi level by ~3 eV. Though, the Mo$4d$-states in 1T-$MoS_2$ form a single wide band, which hosts the Fermi level. In terms of crystal field theory the origin of the latter band can be described as splitting of the Mo$4d$-orbitals of a $O_h$-MoS$_6$ unit into two groups: (1) three degenerated Mo$4d_{xy,yz,xz}$-orbitals populated only by two electrons, and (2) non-occupied Mo$4d_z^2$ and Mo$4d_{x^2-y^2}$ levels (Fig. 6).

The incomplete occupation of Mo$4d_{xy,yz,xz}$-orbitals in 1T-MoS$_2$ leads to the metallic ground state but it also decreases the stability of this MoS$_2$ allotrope. Therefore, by doping the 1T-MoS$_2$ lattice with a donor atom, the additional electrons occupy the Mo$4d_{xy,yz,xz}$-orbitals and increase the stability of the 1T phase. On the contrary, when such doping occurs in the semiconducting 2H-MoS$_2$ allotrope, the electrons which are donated to the Mo$4d_{xy,yz,xz}$ orbitals and to the Mo$4d_{x^2-y^2}$ orbitals result in the metallic-like character of the electronic structure and it causes destabilization of the lattice. Similar analysis of the population of Mo$4d$-orbitals in both MoS$_2$ allotropes can explain the transition from 2H-MS$_2$ lattice to 1T-MS$_2$ lattice in MoS$_2$ and WS$_2$ when intercalated by the electron-donating atoms of alkali-metals [5-7]. Moreover, it suggests that a similar transition and appearance of the 1T-phase may happen as a result of substitutional doping by atoms of $d$-elements which can serve as electron donors. For example, atoms such as Re, Tc, Mn, which have more valence electrons, than Mo or W may act as donors. Our further DFTB calculations for various polytypes of MoS$_2$ doped with different amount of Re atoms support this assumption.

### 3.3. Relative stability of 1T-, 2H- and mixed (1T,2H)-Mo$_{1-x}$Re$_x$S$_2$ allotropes

As the first step the study of the substitution of Mo atoms by Re atoms within 1T- and 2H-MoS$_2$ polytypes was performed by varying the Re content ($x$) and the arrangement of doping atoms. The consideration of low Re concentrations requires the use of supercells of corresponding single- and multi-walled MoS$_2$ nanotubes, which would contain a too large number of atoms. There are also too many possible variants of substitutions for the Re atoms. Thus, in further study we have focused our attention on a more simple case of single and double layers. As the most limiting variants of rhenium distribution, two variants were taken into account: when the Re atoms segregate to make an island of ReS$_2$ phase embedded within the MoS$_2$ lattice and when the Re atoms are scattered randomly within the sublattice of the metal atoms. The relative stability of any doped Mo$_{1-x}$Re$_x$S$_2$ ($x$ = 0.0…1.0) monolayer was characterized as the energy difference $\Delta E$ compared to the 2H-Mo$_{1-x}$Re$_x$S$_2$ monolayer with random distribution of Re atoms.

The calculated values of $\Delta E$ are plotted in Fig. 7. An island-like substitutional Re doping is slightly more favored than a random distribution of Re atoms over a wide range of $x$, for both kinds of MoS$_2$ polytypes. Nevertheless, the maximal energy difference due to the various arrangement of Re atoms at any Re content does not exceed 0.02 eV/atom, which is much smaller relative to the energy differences between the different polytypes. It can be seen that all 1T-Mo$_{1-x}$Re$_x$S$_2$ phases are still clearly higher in energy, than their corresponding 2H-Mo$_{1-x}$Re$_x$S$_2$ phases (Fig. 7). However, the doping may essentially change the energy difference ($\Delta E$) between 1T- and 2H-phase. While for pure MoS$_2$ phases $\Delta E$ is 0.23 eV/atom, it decreases at higher $x$ and adopts its minimal value at $x$ = 0.6, which is only 0.03 eV/atom.

Thus, the calculations show, that the competitive existence of 1T-Mo$_{1-x}$Re$_x$S$_2$ phase could be achieved in practice at a finite temperature and high Re concentration, when the impurity atoms donate the electrons to the Mo sublattice of hosting MoS$_2$ monolayer. However, for the case of a multilayered crystal or multiwalled nanotube a question may appear, if the donor electrons can be adopted from the Re atoms hosted in a neighboring MoS$_2$ layer?

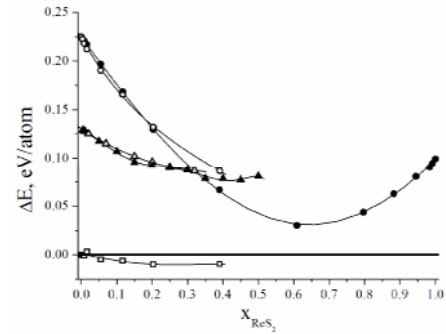

Fig. 7. Relative energies of Mo$_{1-x}$Re$_x$S$_2$ layers depending on the Re content. All energies are given relative to 2H-Mo$_{1-x}$Re$_x$S$_2$ monolayer with random Re distribution. □ – energies of 2H-Mo$_{1-x}$Re$_x$S$_2$ monolayer with "cluster" of Re atoms, ● - energies of 1T-Mo$_{1-x}$Re$_x$S$_2$ monolayer with random Re distribution, ○ - energies of 1T-Mo$_{1-x}$Re$_x$S$_2$ monolayer with "cluster" of Re atoms, ▲ – energies of mixed (1T-MoS$_2$, 2H-Mo$_{1-x}$Re$_x$S$_2$) bilayer with random Re distribution, Δ - energies of (1T-MoS$_2$, 2H-Mo$_{1-x}$Re$_x$S$_2$) bilayer with "cluster" of Re atoms.

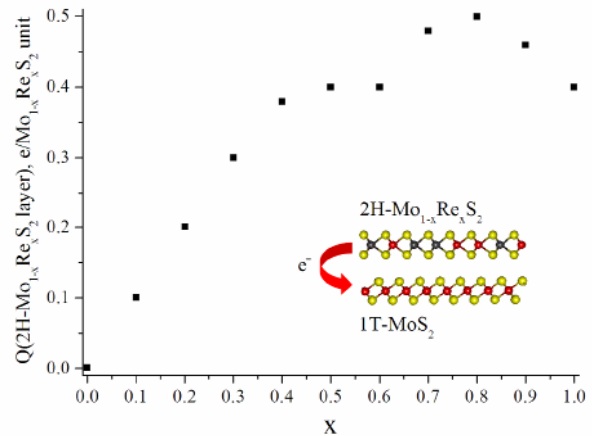

Fig. 8. Charge of 2H-Mo$_{1-x}$Re$_x$S$_2$ layer within (1T-MoS$_2$, 2H-Mo$_{1-x}$Re$_x$S$_2$) bilayer as function of the Re content

To examine this problem similar energy calculations were performed on a MoS$_2$ bilayer composed of adjacent 1T-MoS$_2$ and 2H-Mo$_{1-x}$Re$_x$S$_2$ monolayers (Fig. 7). As it was expected, the relative energy of such mixed (1T,2H)-allotrope for undoped MoS$_2$ is almost the average between the energies of separated 1T- and 2H-MoS$_2$ allotropes. A small deviation from the ideal $\Delta E$ value can be explained by the limitations of the model in describing the superstructure of (1T,2H)-MoS$_2$ composed of two slightly mismatched lattices of 1T- and 2H-MoS$_2$ monolayers. Under doping of the 2H-MoS$_2$ layer, the relative energy $\Delta E$ of mixed (1T-MoS$_2$, 2H-Mo$_{1-x}$Re$_x$S$_2$)-bilayer decreases with increasing of Re content, and it is lower than the $\Delta E$ for the simple 1T-Mo$_{1-x}$Re$_x$S$_2$ allotrope until the doping content reaches $x$=0.35. Noticeably, this correlation between $\Delta E$ for different variants of Re distribution points out that in the mixed and doped (1T,2H)-allotrope the Re atoms should play the same role of the electron donors as in the 1T-Mo$_{1-x}$Re$_x$S$_2$ phase. An analysis of the charge distribution supports this guess (Fig. 8). For a (1T-MoS$_2$, 2H-Mo$_{1-x}$Re$_x$S$_2$)-bilayer an almost linear increase of an electron transfer from the Re-doped layer to the pure MoS$_2$ layer can be seen until $x \approx 0.4$.

## 4. Conclusions

Our work evidences that the occurrence of a 1T-phase revealed by HRTEM methods within disulfide WS$_2$ nanotubes should be stimulated using substitutional doping of Re atoms, which destabilizes the 2H-phase. The Re impurity atoms within the lattice of WS$_2$ or related MoS$_2$ serve as electron-donors, when they are placed within the same layer of the 1T-allotrope as well as placed in the neighboring layer of the 2H-allotrope. Due to equal energy costs, there is no essential preference among both these variants. The distribution of Re atoms within a multiwalled sulfide nanotube should be determined by the fabrication route. For example, in nanotubes fabricated using the gas-phase sulfurization of a mixture of volatile Re and W or Mo chlorides [32], where the reactants themselves already contain Re atoms, Re atoms may be distributed within all the volume of the walls. In contrast, treatment of the already prepared MoS$_2$ or WS$_2$ sulfide nanotubes in the vapor of Re halides (as in the current work) might induce the diffusion of Re atoms only into the lattice of the surface layers.

MoS$_2$ and WS$_2$ are one of the most exploited compounds among transition metal sulfides, which play a major role in the catalytic refinement of petroleum oils and are the main components of technical lubricants in aerospace industry [14,33]. The results of our calculations indicate that Re doping of these nanostructured sulfides could be important in an improvement of both technological applications. It is revealed that the formation of 1T-allotrope as the outer layers of sulfide nanotubes may be induced by the electron-donation from the neighboring doped layers of 2H-phase. This charge transfer reflects the rise of a surface dipole, which points along the normals of layers. Such an enlarged electronic density on the outer walls of a nanotube can be associated with their higher reactivity. This finding opens up a perspective for the Re-doped WS$_2$ and MoS$_2$ nanotubes as superior catalysts similar to the sulfide nanoplatelets [14,17,30].

Numerous former experiments have demonstrated that the quasi-spherical or tubular morphology of 2H-WS$_2$ or 2H-MoS$_2$ nanoparticles, provides a lower density of dangling bonds in comparison to simply nanosized particles of the bulk material and allows a significant increase of the tribological and anti-wear characteristics [33]. The calculations have shown that the strain–stress relationship of the multi-walled sulfide particles is determined by the smallest, innermost wall [34]. Under a load during wear the particles are being broken from the inside to the outside until a bundle of the nanostripes occurs, which are always parallel to the working surfaces and form nanocoating [34]. Obviously even these superior tribological characteristics of MoS$_2$ and WS$_2$ fullerene-like particles and nanotubes can be improved further using Re doping. The present calculations reveal a higher (comparing with 2H-MoS$_2$) strain and, following, a lower mechanical resistance of 1T-MoS$_2$ shells. When the 1T phase occurs at the outer surface of doped multiwalled sulfide nanotubes, a lower wear coefficient can be achieved due to the more premature break and exfoliation of the outer shells. These may serve later, after the adhesion on the working surfaces, as a smear undercoat for the rest of the multiwalled particles, and provides the better bearing effect. Further experimental as well as theoretical studies should verify these preliminary suggestions.

## Acknowledgments

The support of the ERC project INTIF 226639 is gratefully acknowledged. A.E. thanks grant RFBR 11-03-00156-a.

## Notes

[a] *Physical Chemistry Department, Technical University Dresden, Bergstr. 66b, 01062 Dresden, Germany. Fax: +49 (0) 351 463 35953; Tel: (+49) (351) 463 37637; E-mail: enyashin@ihim.uran.ru; gotthard.seifert@chemie.tu-dresden.de*
[b] *Institute of Solid State Chemistry UB RAS, Pervomayskaya Str. 91, 620990 Ekaterinburg, Russia. Fax: +7 343 374 4495; E-mail: enyashin@ihim.uran.ru*
[c] *Materials and Interfaces Departament, Weizmann Institute of Science, 76100 Rehovot, Israel. E-mail: reshef.tenne@weizmann.ac.il*
[d] *Peter Grünberg Institute, Ernst Ruska-Centre for Microscopy and Spectroscopy with Electrons, Research Centre Juelich, 52425 Juelich, Germany. Fax: +49 (0) 246 161 6444; E-mail: m.bar-sadan@fz-juelich.de; l.houben@fz-juelich.de*
[e] *School of Physics and CRANN, Trinity College, Dublin 2, Ireland.*